\documentclass[10pt, sigconf, authorversion]{acmart}

\usepackage[acronym, nowarn]{glossaries}
\makeglossaries

\newacronym{manet}{\textsc{MANET}}{mobile ad hoc network}
\newacronym{dtn}{\textsc{DTN}}{delay-tolerant network}
\newacronym{d2d}{\textsc{D2D}}{device-to-device}
\newacronym{los}{\textsc{LoS}}{Line-of-Sight}
\newacronym{gps}{\textsc{GPS}}{Global Positioning System}
\newacronym{one}{\textsc{ONE}}{Opportunistic Network Simulator}
\newacronym{smarter}{\textsc{SMARTER}}{\textbf{Smart}phone-based Communication Networks for \textbf{E}mergency \textbf{R}esponse}
\usepackage{booktabs} % For formal tables
\usepackage{balance}
\usepackage{todonotes}
\usepackage{natbib}
\usepackage{xspace}
\usepackage{tabularx}
\usepackage{hhline}
\usepackage{hyperref}
\usepackage[caption=false,font=footnotesize]{subfig}
\usepackage{dblfloatfix}
\usepackage[babel=true]{csquotes}

\newcommand{\eg}{e.\,g.\xspace}

\newcommand\doublerule{\hline\hline}
\newcommand\doubleheavyrule{\hline\specialrule{\heavyrulewidth}{\doublerulesep}{0.0em}}
\newcommand{\textbfit}[1]{\textbf{\textit{#1}}}

\usepackage{hyperref}
\usepackage[hyphenbreaks]{breakurl}

\begin{document}
\copyrightyear{2018} 
\acmYear{2018} 
\setcopyright{acmcopyright}
\acmConference[CHANTS '18]{13th Workshop on Challenged Networks}{October 29, 2018}{New Delhi, India}
\acmBooktitle{13th Workshop on Challenged Networks (CHANTS '18), October 29, 2018, New Delhi, India}
\acmPrice{15.00}
\acmDOI{10.1145/3264844.3264845}
\acmISBN{978-1-4503-5926-9/18/10}
\fancyhead{}
\title[Field Test of a Smartphone-based Communication Network for Emergency Response]{Conducting a Large-scale Field Test of a Smartphone-based Communication Network for Emergency Response}
\author{Flor ~\'{A}lvarez}
\authornote{Authors contributed equally to the paper.}
\affiliation{%
%  \department{Secure Mobile Networking Lab}
 \department{SEEMOO}
  \city{TU Darmstadt}
  \country{Germany}
  %\postcode{D-64653}
}
\email{falvarez@seemoo.tu-darmstadt.de}
\author{Lars Almon}
\authornotemark[1]
\affiliation{%
%  \department{Secure Mobile Networking Lab}
 \department{SEEMOO}
  \city{TU Darmstadt}
  \country{Germany}
  %\postcode{D-64653}
}
\email{lalmon@seemoo.tu-darmstadt.de}
\author{Patrick Lieser}
\authornotemark[1]
\affiliation{%
  %\department{Multimedia Communications Lab}
   \department{KOM}
 \city{TU Darmstadt}
  \country{Germany}
  %\postcode{D-64289}
}
\email{patrick.lieser@kom.tu-darmstadt.de}
\author{Tobias Meuser}
\authornotemark[1]
\affiliation{%
 % \department{Multimedia Communications Lab}
    \department{KOM}
 \city{TU Darmstadt}
  \country{Germany}
  %\postcode{D-64289}
}
\email{tobias.meuser@kom.tu-darmstadt.de}
\author{Yannick Dylla}
\affiliation{%
  %\department{Secure Mobile Networking Lab}
   \department{SEEMOO}
  \city{TU Darmstadt}
  \country{Germany}
  %\postcode{D-64653}
}
\email{ydylla@seemoo.tu-darmstadt.de}
\author{Bj\"orn Richerzhagen}
\orcid{0000-0002-9584-8682}
\affiliation{%
 % \department{Multimedia Communications Lab}
   \department{KOM}
  \city{TU Darmstadt}
  \country{Germany}
 % \postcode{D-64289}
}
\email{richerzhagen@kom.tu-darmstadt.de}

\author{Matthias Hollick}
\affiliation{%
 % \department{Secure Mobile Networking Lab}
  \department{SEEMOO}
  \city{TU Darmstadt}
  \country{Germany}
  %\postcode{D-64653}
}
\email{mhollick@seemoo.tu-darmstadt.de}

\author{Ralf Steinmetz}
\affiliation{%
	%\department{Multimedia Communications Lab}
	   \department{KOM}
	\city{TU Darmstadt}
	\country{Germany}
	%\postcode{D-64289}
}
\email{ralf.steinmetz@kom.tu-darmstadt.de}

% The default list of authors is too long for headers.
\renewcommand{\shortauthors}{F. Alvarez, L. Almon, P. Lieser, T. Meuser, et al.}

\begin{abstract}
%!TEX root = ../paper_traces.tex

Smartphone-based communication networks form a basis for services in emergency response scenarios, where communication infrastructure is impaired or overloaded.
Still, their design and evaluation are largely based on simulations that rely on generic mobility models and weak assumptions regarding user behavior.
For a realistic assessment, scenario-specific models are essential.
To this end, we conducted a large-scale field test of a set of emergency services that relied solely on ad hoc communication.
Over the course of one day, we gathered data from smartphones distributed to 125 participants in a scripted disaster event. 
In this paper, we present the scenario, measurement methodology, and a first analysis of the data.
Our work provides the first trace combining user interaction, mobility, and additional sensor readings of a large-scale emergency response scenario, facilitating future research.

\end{abstract}

%
% The code below should be generated by the tool at
% http://dl.acm.org/ccs.cfm
% Please copy and paste the code instead of the example below.
%
\begin{CCSXML}
<ccs2012>
<concept>
<concept_id>10003033.10003079.10003082</concept_id>
<concept_desc>Networks~Network experimentation</concept_desc>
<concept_significance>500</concept_significance>
</concept>
<concept>
<concept_id>10003033.10003106.10010582.10011668</concept_id>
<concept_desc>Networks~Mobile ad hoc networks</concept_desc>
<concept_significance>500</concept_significance>
</concept>
<concept>
<concept_id>10003033.10003083.10003094</concept_id>
<concept_desc>Networks~Network dynamics</concept_desc>
<concept_significance>300</concept_significance>
</concept>
<concept>
<concept_id>10003033.10003083.10003097</concept_id>
<concept_desc>Networks~Network mobility</concept_desc>
<concept_significance>300</concept_significance>
</concept>
</ccs2012>
\end{CCSXML}

\ccsdesc[500]{Networks~Network experimentation}
\ccsdesc[500]{Networks~Mobile ad hoc networks}
\ccsdesc[300]{Networks~Network dynamics}
\ccsdesc[300]{Networks~Network mobility}
\settopmatter{authorsperrow=4}
\maketitle

\keywords{field test; emergency response; smartphone-based communication}

%!TEX root = ../paper_traces.tex
\section{Introduction}
\label{sec:introduction}
Recent disasters such as the hurricanes Maria, Jose, and Harvey in 2017
demonstrated the challenges of disaster relief efforts.
Especially when disasters strike in urban environments, working
information and communication infrastructure is key for emergency
response.
However, infrastructure is either severely impaired or overloaded after such disasters.
To this end, resilient infrastructure independent ad hoc
communication services can be established by facilitating the smartphones of civilians and responders.
Using these communication services, a number of applications for medical care~\cite{malan2004codeblue} and coordination of those affected~\cite{al2014help} can be realized.

To asses the effectiveness of the proposed applications and to improve them upon the lessons learned, they would be ideally tested during a real crisis situation.
This is infeasible on a regular and planned basis, leading to the utilization of simulations models during the design phase of applications and services.
Most of the time, the evaluation of such applications is based on simulation models, trying to mimic realistic user behavior and environment characteristics for post-disaster scenarios.
These models are either (i) solely based on the analysis of tactical issues of civil protection and input from FEMA or other organizations, (ii) relying on traces gathered in everyday life, e.g., on campuses, during conferences, or in office buildings \cite{parris2014facebook}, or (iii) only considering behavior of professional disaster relief personnel~\cite{Stute2017}.
They all miss important characteristics of real-world human behavior---especially of civilians.
This severely limits their applicability to evaluate the aforementioned services relying on ad hoc networks.
As surveyed in \cite{aschenbruck2011trace}, there is a plethora of trace-based movement models based on real human movement records. 
However, most of them cover everyday movement patterns.

To address this issue, we conducted a field test mimicking a post-disaster situation as realistic as possible. 
We recorded the user behavior and user interaction of 125 participants. 
Additionally, we conducted a questionnaire after the field test to asses the subjective experience when interacting with specific disaster services.
This is the first work describing a disaster scenario and measuring the behavior of those affected on a sufficiently detailed level to be used as a foundation for simulations of present and future disaster communication services.\footnote{Smarter-dataset [Online]. Available: https://seemoo.de/smarterfield-test}
This paper is structured as follows. 
We provide a description of the field test setup in Section \ref{sec:scenario}.
Section \ref{sec:analysis} provides a first analysis and discussion of the collected data, highlighting scenario-specific interaction and movement behavior of participants.
We discuss the resulting implications and future work in Section \ref{sec:conclusion}.
%!TEX root = ../paper_traces.tex
\section{Field Test}
\label{sec:scenario}
The field test took place in September 2017 at the military training area \textit{Senne} near Paderborn in Germany in conjunction with experts from the German Federal Office of Civil Protection and Disaster Assistance (BBK), the German Federal Agency for Technical Relief (THW), local fire departments, and other NGOs. 
\begin{figure}[H]
	\centering
		\includegraphics[width=0.95\columnwidth]{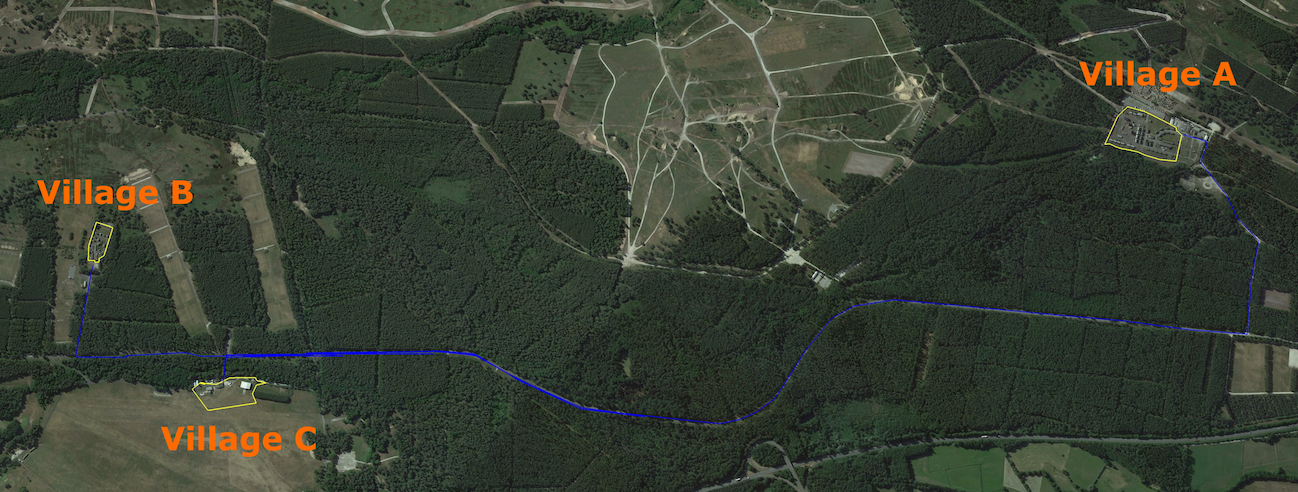}
		\vspace{-0.15cm}
	\caption{Layout of the military training area Senne}
		\vspace{-0.15cm}
	\label{fig:layoutSenne}
\end{figure}

\autoref{fig:layoutSenne} shows the layout of the field test area containing three villages (A, B, C) equipped with brick buildings. 
The linear distance between villages B and C is $700\,$m and between A and B is $4\,$km.
125 volunteers participated in the test between 09:30 and 16:30.\footnote{Readers can get an impression of the field test by visiting: \url{www.youtube.com/watch?v=Hb8mgVJHrs0}.}
Participants had to find family members, help, and resources after a complete breakdown of the communication infrastructure caused by a grid blackout.
To evaluate behavior in stressful situations, two fictive events took place during the field test, involving professional actors.
In village A, a lightning strike hit a gas station at 13:00 and injured a couple of people with the need for immediate help and shelter.
In village B, hazardous substances were released at 14:30 after cooling at a chemical plant failed, requiring immediate evacuation.
Actors further increased distress by, \eg, playing a mother desperately searching for her child.
As motivated, the main goal of the field test was the evaluation of a smartphone-based ad
hoc network supporting a set of emergency services (\eg, \textit{SOS Emergency
	Messages}, \textit{Resource Market Registry}) as
described in \cite{lieser2017}. 
In addition to technical insights into the underlying ad hoc network, we also addressed the usability and utilization of the proposed services in a realistic scenario.
The services were implemented within an Android application\footnote{\url{http://smarter-projekt.de/demonstrator/}}, and direct communication between nearby devices relied on IBR-DTN~\cite{Morgenroth:2012:BPI:2348543.2348606}, using Wi-Fi in ad hoc mode.
We chose Google Nexus 5, 6P and Samsung Galaxy S6 devices for the field test as we had already experience with them for enabling in the ad hoc mode~\cite{nexmon}.
We log all user interaction with our app. We also used a custom logging framework to capture sensor data and network statistics.
All measurements were tagged with the device-specific unique DTN-ID provided by IBR-DTN.
\vspace{-0.2cm}
\subsection{Setup}
At the beginning of the field test, participants received a smartphone and a portfolio with information about their character.
The character was completely fictitious to protect the privacy of the participants. 
The participants were distributed over the three villages. 
The portfolio contained the home address (village), age, and family relations of the respective character.
Additionally, tasks like search for your family members, meet at the home address, or search for specific resources such as water or medical supply were stated.
Each participant started with at least three resources with additional resources being deployed throughout the field test area as a motivation to utilize the \emph{Resource Market Registry} of the application.
The application running on each device was pre-configured with a personalized address book containing only contacts according to the portfolio of the respective character.
\vspace{-0.25cm}
\subsection{Data Collection}
During the whole field test, we recorded sensor, network, and application-related data.
To compensate the increased energy consumption, each participant received a battery pack with sufficient energy for the duration of the field test.
Sensor data was recorded on average every second and saved in a local SQLite database. 
We recorded GPS locations, accelerometer readings, brightness, air pressure, and gyroscope readings.
Our previous research shows, that the data gathered from this set of sensors supplies sufficient information to recognize a person activity, as well as to differentiate if a person performs a disaster related activity such as crawling on the floor or walking with an injured leg \cite{liesersituation}.
The brightness sensor can be used to determine if the phone is in the pocket or held in the hand of a user.
The sensor data can be used for a number of future research directions, as discussed in Section~\ref{sec:conclusion}.

Regarding our prime objective of assessing the performance of the smartphone-based ad hoc network, we logged all network-related information provided by IBR-DTN.
This includes information about locally generated data bundles, transmitted and received bundles, connection events between devices, and discovered peers.
Based on this data, we can assess the store-carry-forward principle of the delay tolerant communication network.

To assess the general utilization and usability of the proposed services, we recorded information related to interaction with the application on each device.
This included tracking each interaction---i.e., screen taps---and each event generated by the application, e.g., incoming notifications or new information being displayed.
All data was stored locally with a timestamp and the device's DTN-ID.

%!TEX root = ../paper_traces.tex
\vspace{-0.20cm}
\section{Analysis of the Dataset}
\label{sec:analysis}
\begin{figure}[ht!]
	\centering
	\vspace{-0.20cm}
	\includegraphics[width=0.50\columnwidth]{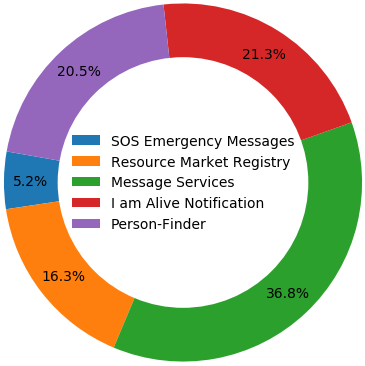}
	\caption{Service Usage}
		\vspace{-0.45cm}
	\label{fig:all-messages}
\end{figure}

We investigate the performance and scalability aspects of the simulated scenario by analyzing delay and hop distribution, number of neighbors, participant speed, and connection data between mobile devices.
To prevent the results from being inconsistent by the fact that the participants were transported in a bus to each village, we have considered only the data collected between 10:30 and 15:30 for our analysis. 
Due to various problems: hardware (SD card to slow, etc.), software (app malfunctions, etc.), user device handling, and a lost device, we could not gather a complete dataset.
Out of the 125 devices, 119 contributed to the network and app dataset and 96 were used to built the GPS traces. 
\vspace{-0.1cm}
\paragraph{Data validation and cleaning}
After joining the collected data into one database, we detected that some devices had more than one unique DTN-ID during the experiment.
To avoid inconsistent analysis, we matched all DTN-IDs to the corresponding device. 
During GPS data analysis we observed a difference between smartphones models: Google Nexus devices provide a consistent GPS data and mostly logged one time per second.
By contrast, the GPS data from Samsung Galaxy devices is irregular and mostly logged up to two times per second.
%\vspace{-0.4cm}
In addition, we have found difference in the timestamp associated of the logs between devices. 
As the devices had neither access to Internet nor connection to any other time synchronization source, it was not possible to have a perfect time synchronization between all devices. 
Because of that, we consider the devices with the most number of connections (from 90 connections) as those 
with the reference time, i.e., we took the timestamp of those as the ground truth and synchronized all other devices based on this information. 
\begin{table}[H]
	\vspace{-0.3cm}
	\renewcommand{\arraystretch}{1.3}
	\caption{Dataset summary}
	\label{tab:data_summary}
	\vspace{-0.38cm}
	\resizebox{0.48\textwidth}{!}{
		\begin{tabular}{lccc}
			\doubleheavyrule
											& \textbfit{Mean}	& \textbfit{Standard deviation}	& \textbfit{Median}\\
			\hline
			Connection distance (m)			&  44.21	&  41.35	&   30.02\\
			Contact duration (s)				& 301.88	& 624.69	&   97.00\\
			Walking distance (km)			&  11.39	&   4.59	&   11.46\\
			Walking speed (km/h)			&   2.14	&   2.85	&    0.72\\
			Number of neighbors (d = 44m)	&   7.20	&   2.78	&    7.00\\
			Message size (byte)				& 290.15	& 568.76	& 1,835\\
			Multicast delay (min)					&  19.89	&  18.33	&   15.22\\
			Multicast propagation (min)			&  26.72	&  19.33	&   27\\
			Cluster coefficient				&   0.31	&   0.05	&    0.30\\
			\doublerule
		\end{tabular}
	}
	\vspace{-0.3cm}
\end{table}
\autoref{tab:data_summary} summarizes the most important information results from the analysis of the sensor and network data.
Additionally, Figure \ref{fig:all-messages} summarizes the emergency services usage in the whole field test. 
\vspace{-0.20cm}
\vspace{-0.35cm}
\subsection{Sensor Data}
In this section we analyze the information about GPS tracks, number of neighbors and walking speed.
\subsubsection{Participants walking speed}
\begin{figure}[H]
	\centering
	\vspace{-0.35cm}
	\includegraphics[width=0.35 \textwidth]{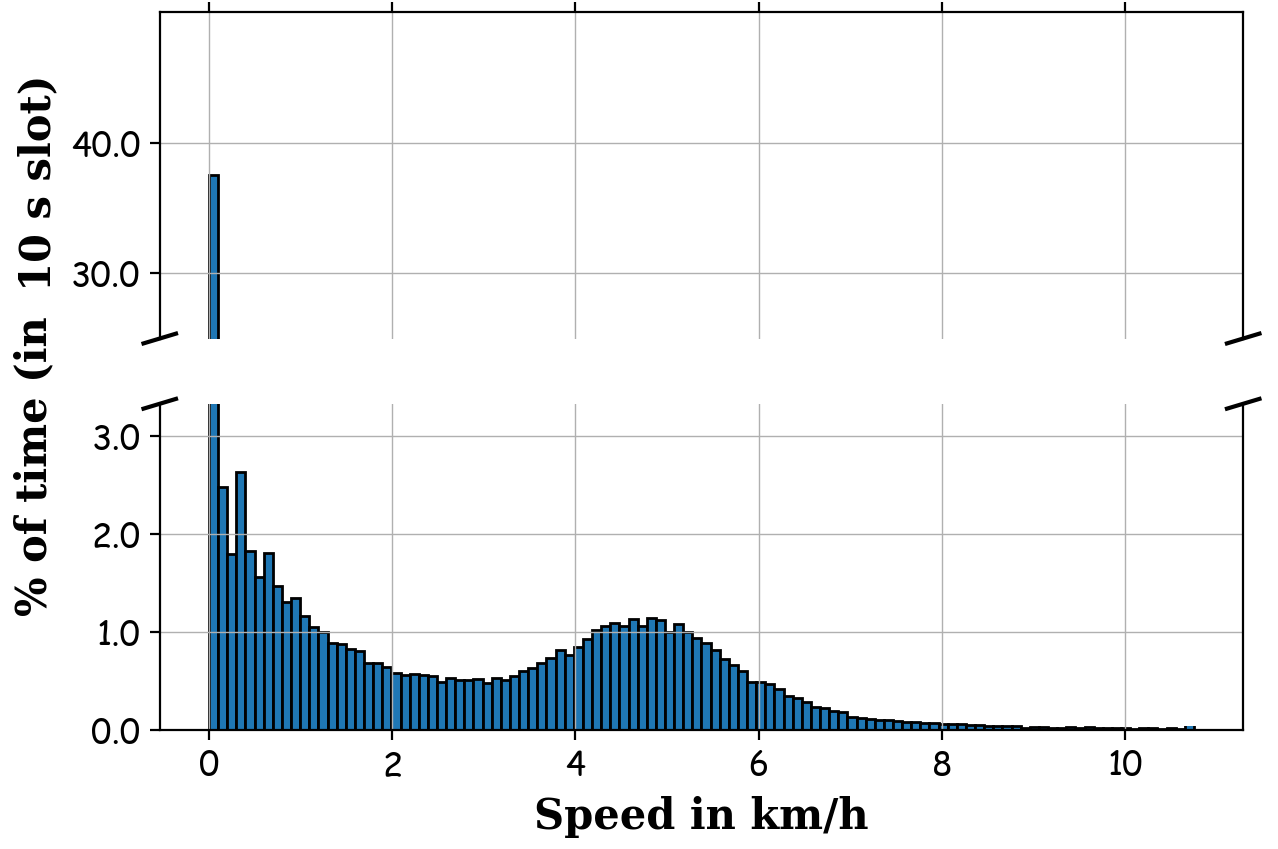}
	\vspace{-0.15cm}
	\caption{Walking speed of the participants}
	\label{fig:speed_hist}
	\vspace{-0.30cm}
\end{figure}

We analyze the participant speed recorded along the field test in Figure \ref{fig:speed_hist}, which confirm previous results 
about the normal person speed with an average of 1.9 km/h \cite{aschenbruck2007modelling}.
We also observed quite static behaviors of participants (around 35 percent of the time), with few peaks corresponding to speeds between 1 km/h and 4-5 km/h.
\vspace{-0.3cm}
\vspace{-0.60cm}
These values are the result of the mobility pattern reproduced by our specific scenario: the static time represents i.e., breaks in each new encounter in order to exchange information and resources.
The peaks are the contribution of the participant movement from a village to another one. 
\vspace{-0.10cm}
\subsubsection{Participants GPS tracks}
\begin{figure*}[t!]
	\centering
	\vspace{-0.35cm}
	\subfloat[]{
		\includegraphics[width=0.275 \linewidth]{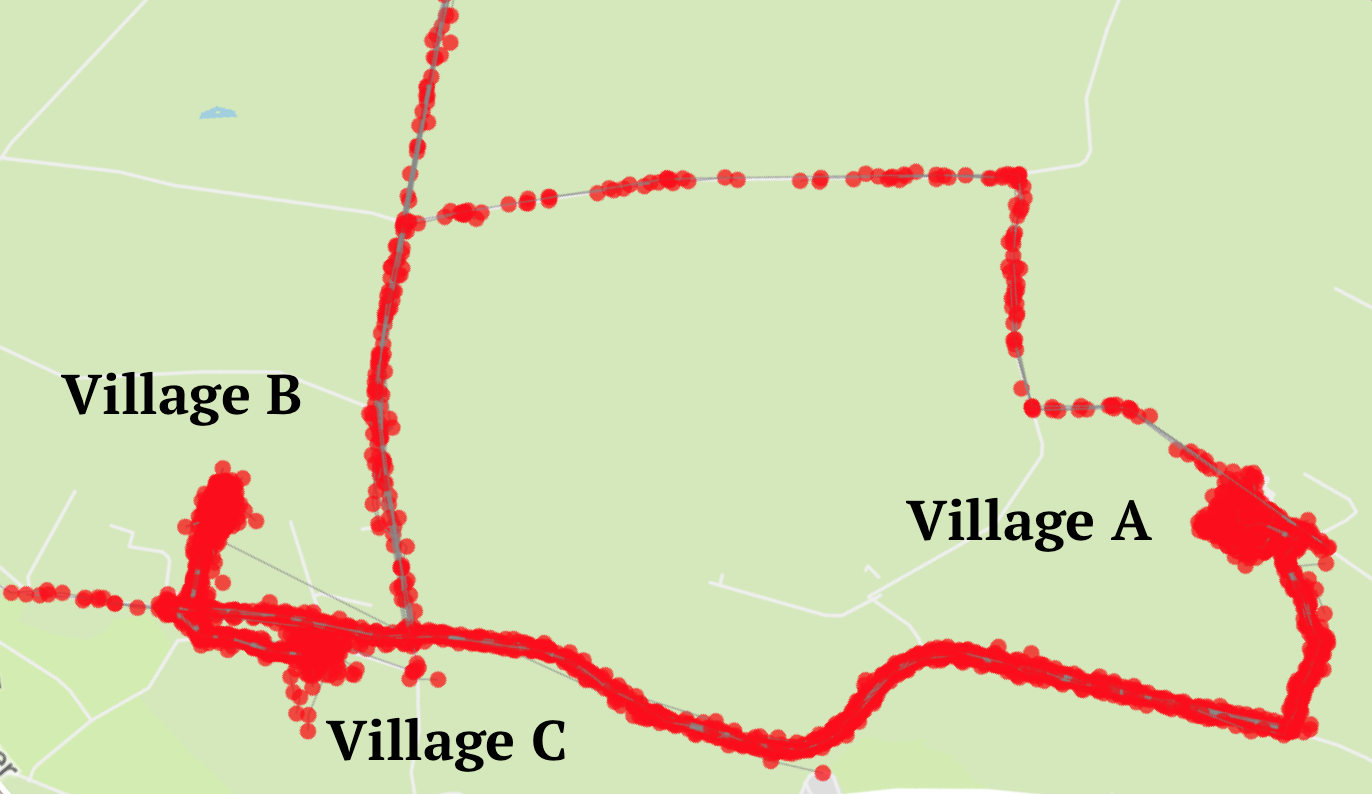}
	\label{fig:gps_track_all}
	}
	\subfloat[]{
		\includegraphics[width=0.225 \linewidth]{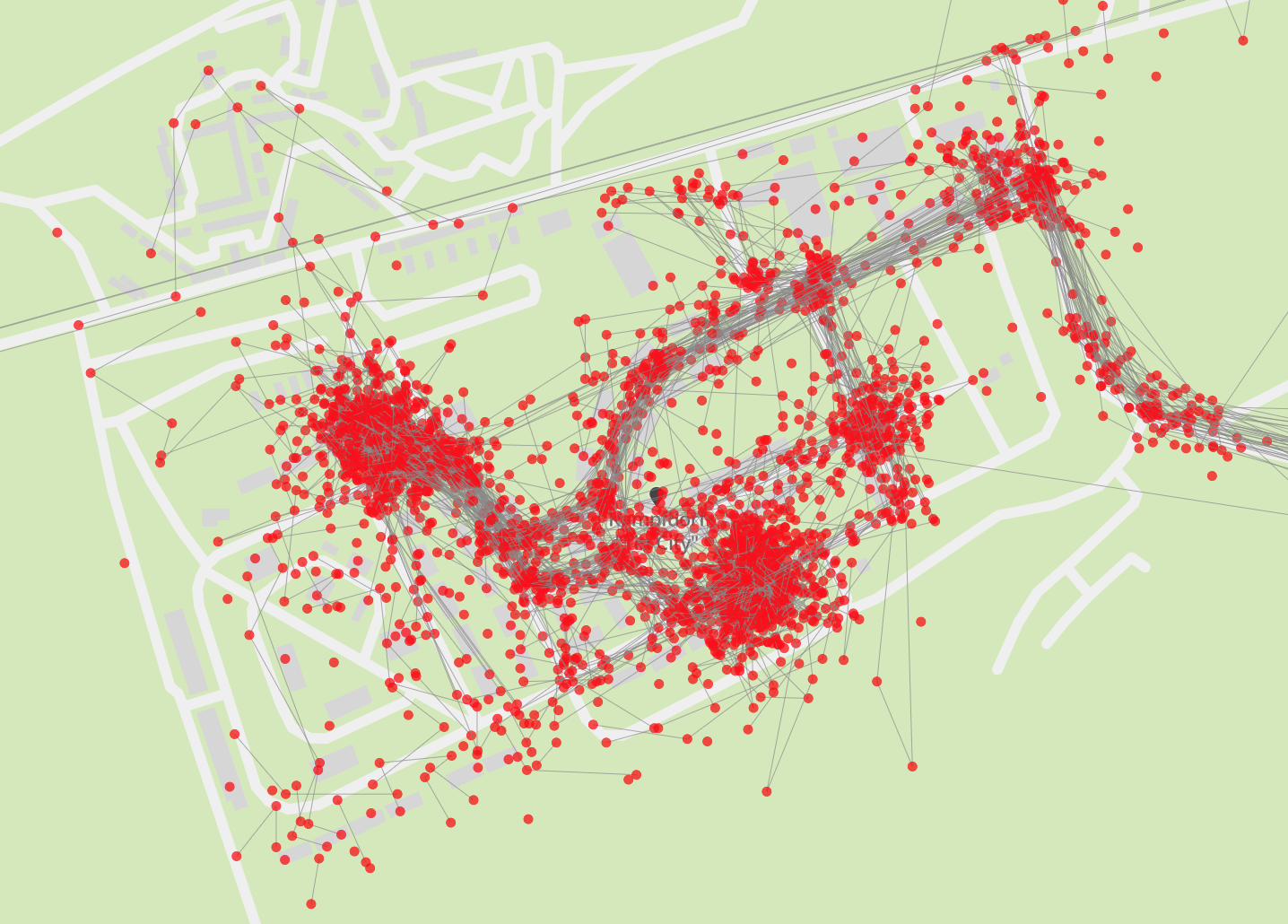}
		\label{fig:gps_track_A}
	}
	\subfloat[]{
		\includegraphics[width=0.215\linewidth]{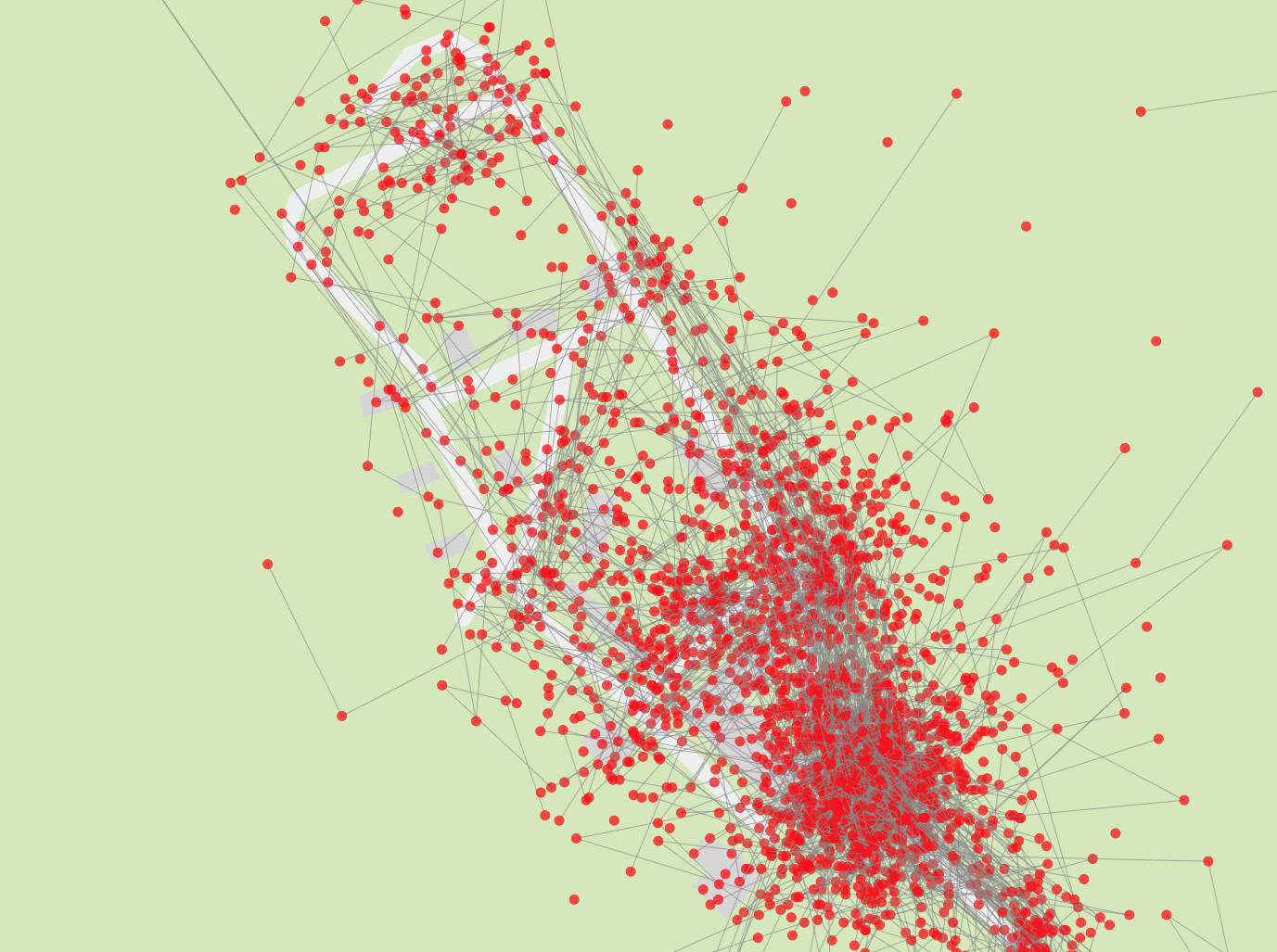}
		\label{fig:gps_track_B}
	}
	\subfloat[]{
		\includegraphics[width=0.215 \linewidth]{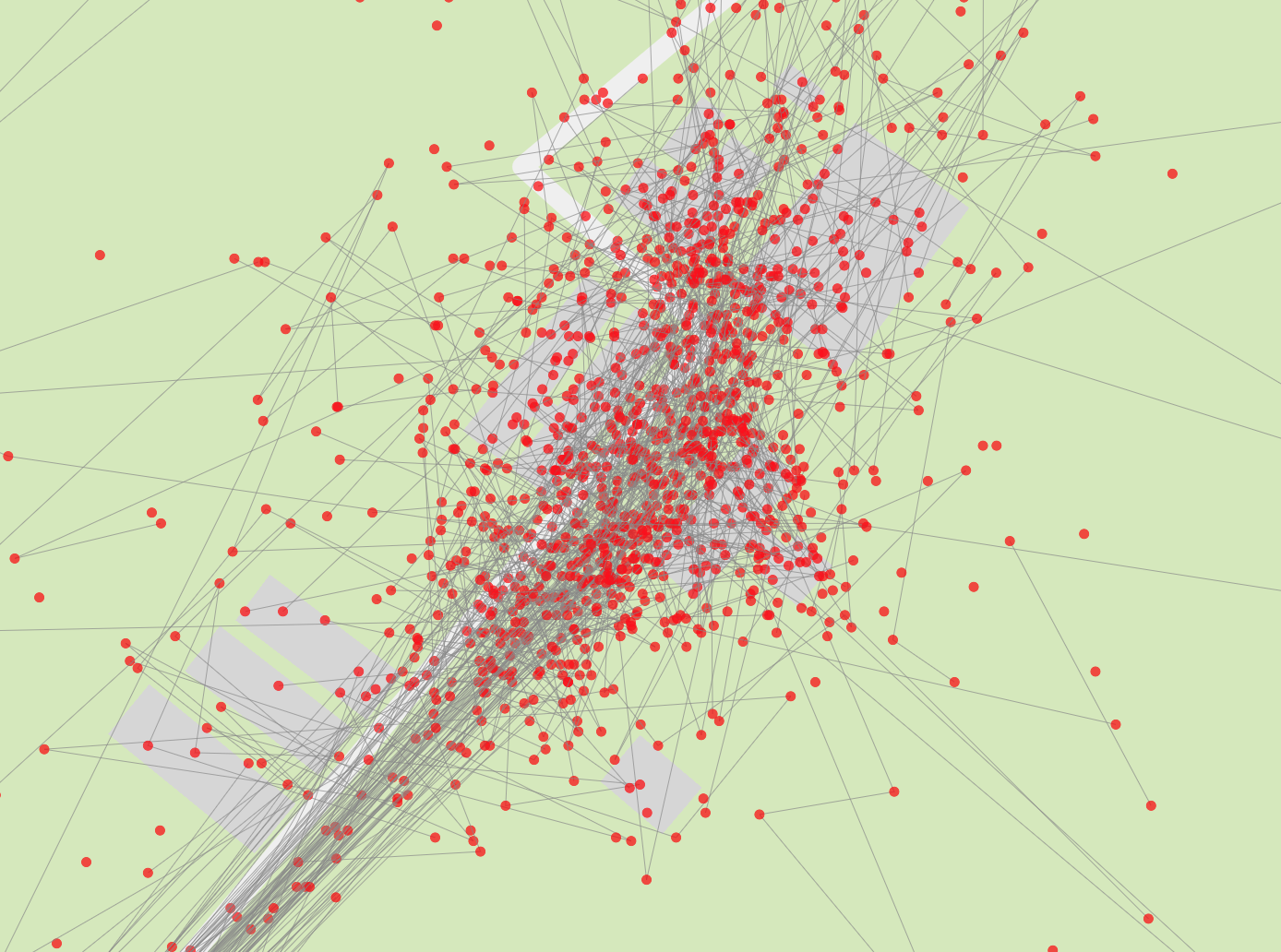}
		\label{fig:gps_track_C}
	}
	\vspace{-0.35cm}
	\caption{GPS Track: (a)  training area Senne, (b) Village A, (c) Village B, and (d) Village C}
	\label{fig:gps_track}
	\vspace{-0.30cm}
\end{figure*}

By using the GPS data, we replicated the movement of each participant throughout the field test as depicted in Figure \ref{fig:gps_track}. 
Most of the participants stayed on the planned route.
However, there were also some users, who used alternative routes to mobilize. 
\vspace{-0.10cm}
\subsubsection{Number of neighbors}
We use the GPS data of each device to quantify the number of neighbors that each participant had throughout the field test. 
For our analysis, we choose three values to set the maximal distance between two devices considered neighbors: 25, 44 and 110 m.
We took these values based on the results from the analysis of the network data as shown in Figure \ref{fig:connection} (b): 
most of the 50 percent up-connection were within approx. 25 m, the mean was around 40 m, and 90 percent of the connections were within 110m. 
\vspace{-0.7cm}
\begin{figure}[H]
	\includegraphics[width=0.7\linewidth]{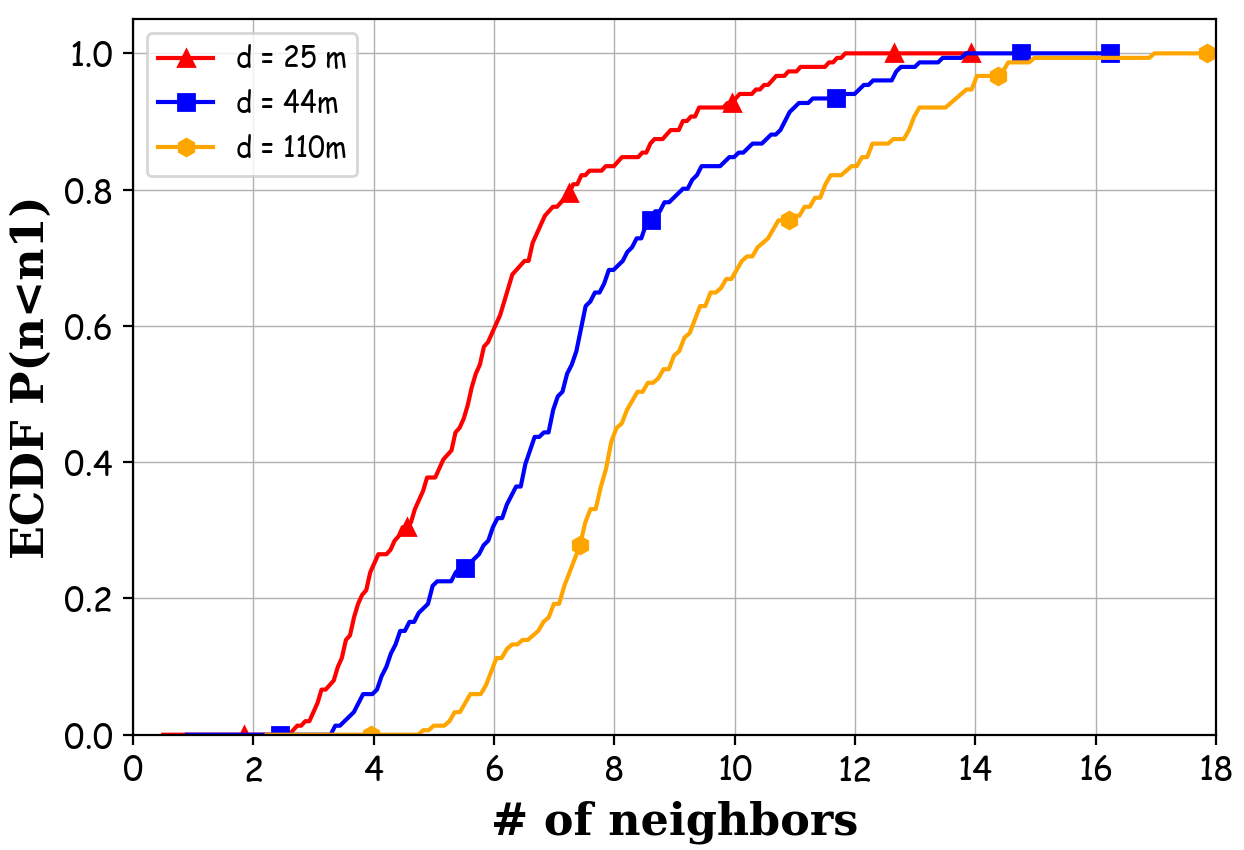}
	\label{fig:ecdf}
	\vspace{-0.35cm}
	\caption{Neighbor aggregated over 2 minutes as ECDF}
	\label{fig:neighbor}
	\vspace{-0.35cm}
\end{figure}
\vspace{-0.3cm}
On average, each participant had between six and eight neighbors, as Figure \ref{fig:neighbor} shows. 
Many of the groups were built upon the relationships between users as described in the portfolio.
But, we also found that participants moved most of the time in small groups, including persons who are not in their family circle. 
\vspace{-0.15cm}
\begin{figure}[H]
	\centering
	\vspace{-0.3cm}
	\includegraphics[width=0.7\linewidth]{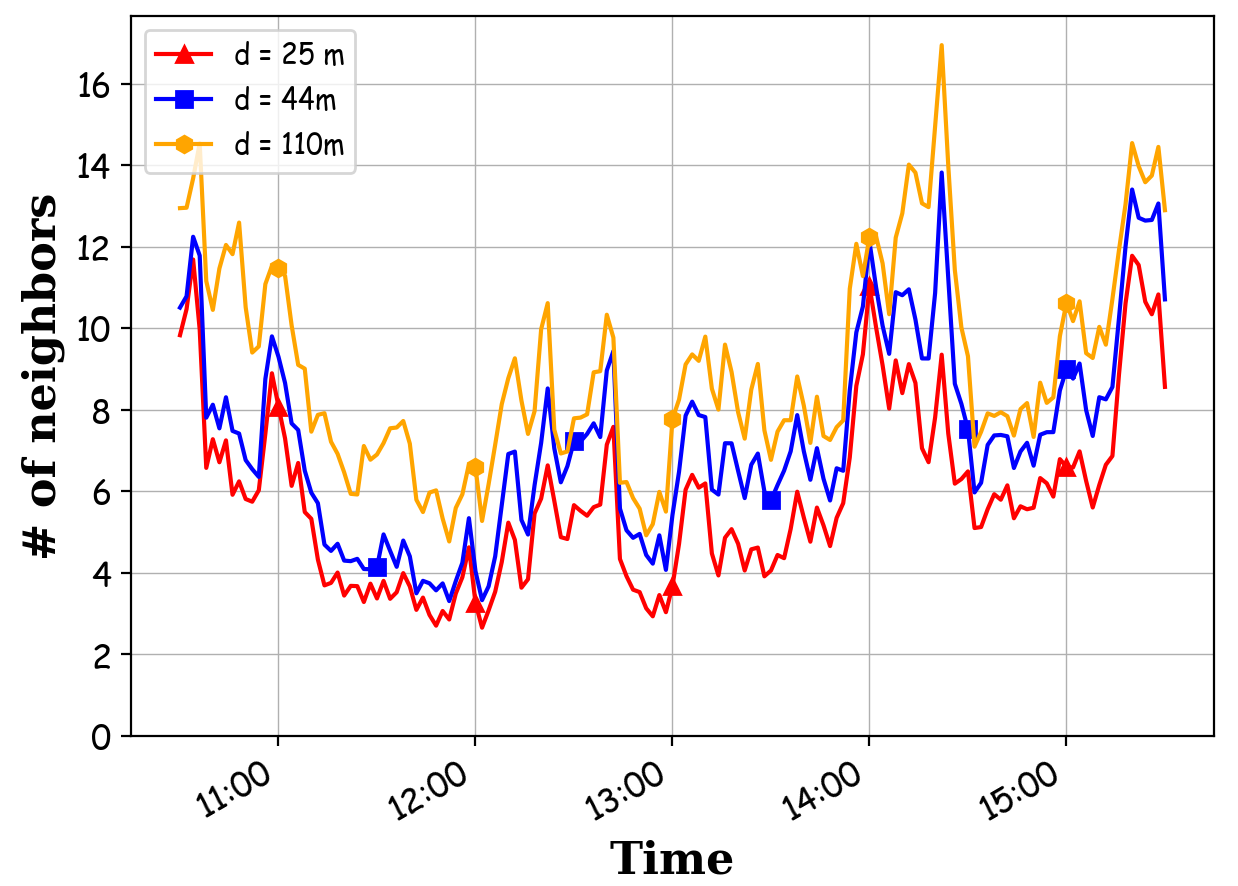}
	\vspace{-0.35cm}
	\caption{Neighbors aggregated over 2 min. over the time}
		\label{fig:neighbor_time}
	\vspace{-0.35cm}
\end{figure}
\vspace{-0.1cm}
Based on Figure \ref{fig:neighbor_time} we recognize additional characteristics of the user behavior in our experiment: 
most of the contacts occurred around 10:30 and between 13:00 and 15:00.
This result is reasonable, since these peaks represent the start of the test as well as our two simulated events. 
Moreover, even in the walking phase most of the device had at least three neighbors. 
\vspace{-0.1cm}
\subsection{Network Data}
\label{sub:networkdata}
\subsubsection{Connection}
\begin{figure*}[t!]
	\centering
	\vspace{-0.45cm}
	\subfloat[]{
		\includegraphics[width=0.30 \linewidth]{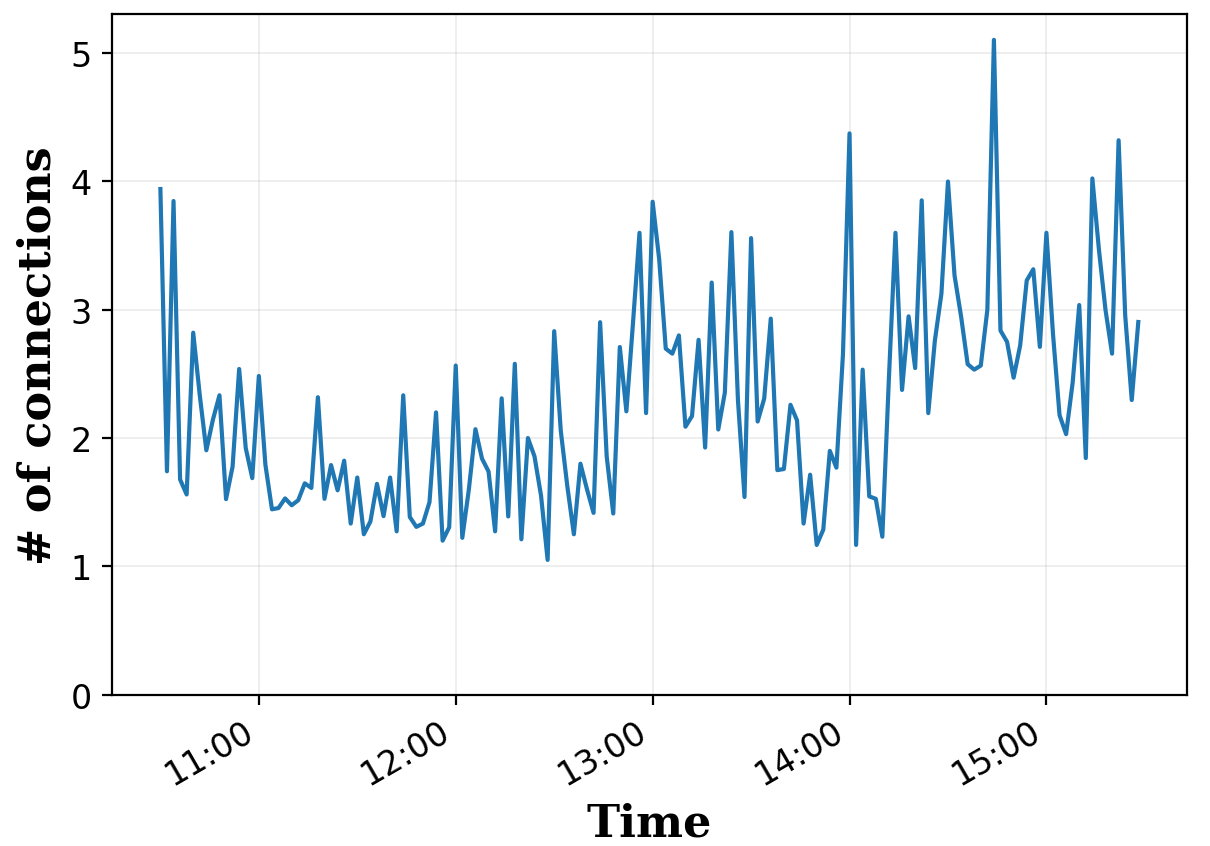}
	\label{fig:connection_count}
	}
	\subfloat[]{
		\includegraphics[width=0.315 \linewidth]{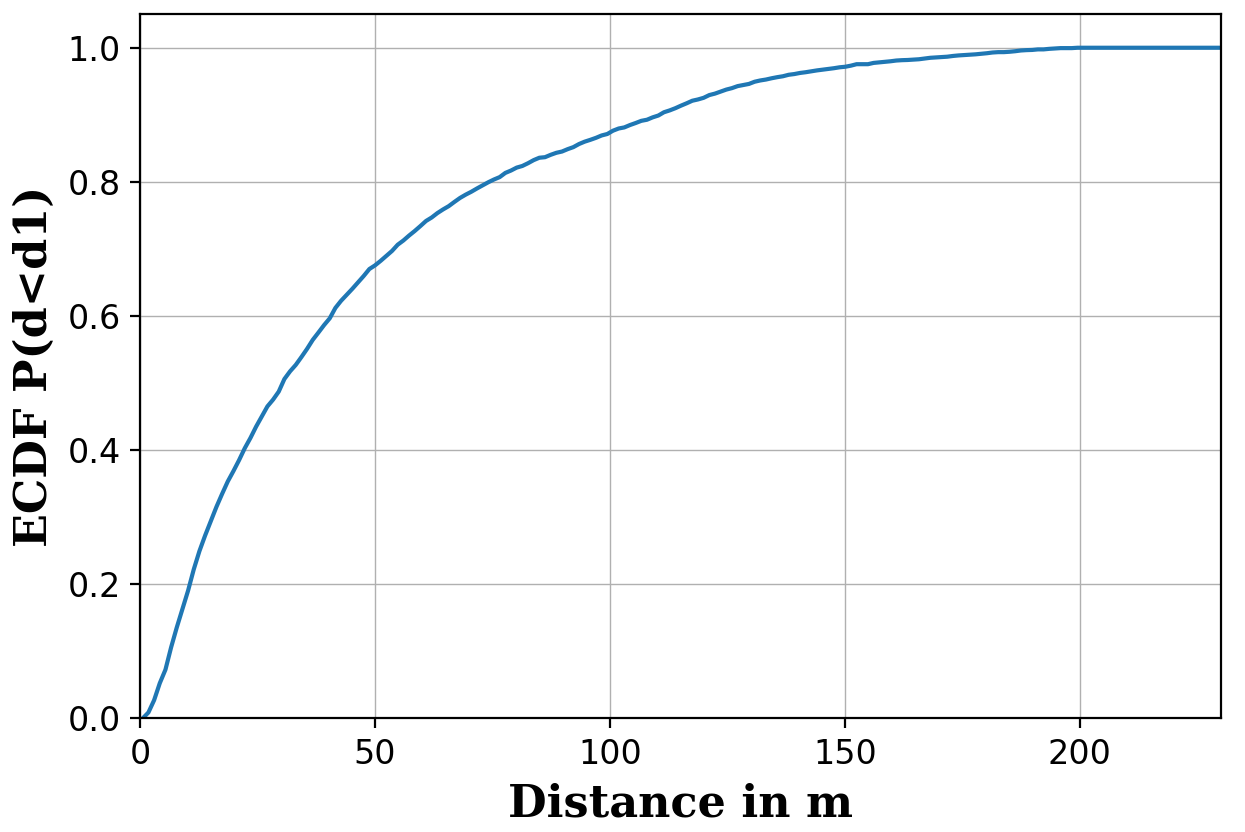}
	\label{fig:ecdf_distance}
	}
	\subfloat[]{
		\includegraphics[width=0.32 \linewidth]{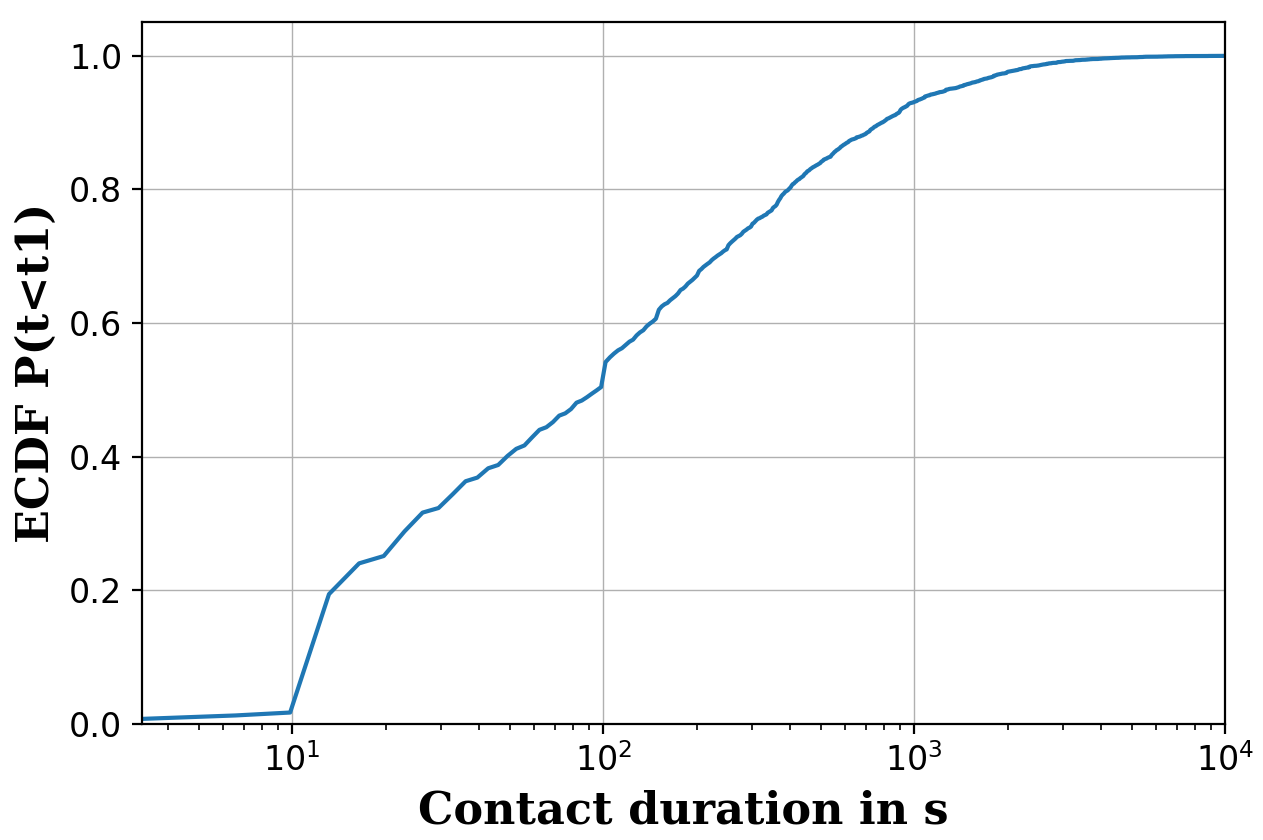}
		\label{fig:ecdf_duration}
	}
	\vspace{-0.35cm}
	\caption{Connection distribution aggregated over 2 minutes: (a) average device connections over the time, (b) ECDF distance, and (c) ECDF duration}
	\label{fig:connection}
	\vspace{-0.35cm}
\end{figure*}
Based on the data, we explored information about the number of connections, connection duration and connection distance of a device pair.
We analyzed the empirical distribution of the connection duration and distance. 
Figure \ref{fig:connection} (a) visualizes the connections distribution over the time.
The observed peaks match with our two simulated events and the end of the field test. 
By comparing this result with the number of neighbors obtained from the GPS data, we can conclude that both distribution present a similar behavior. 
As depicted in Figure \ref{fig:connection} (b), 90 percent of the up-connection were within approx. 110 m.
This value can be considered as the expected in an area where a free \gls{los} is given. 
Yet, connection distances of over 150 m where possible too.
Moreover, we also visualize in Figure \ref{fig:connection} (c) the empirical cumulative distribution function using a log scale for the x-axis of the duration of a connection between two devices.
Interestingly, we found that most of the connection had a duration time of 100 seconds. 
This information can impact assumptions and decisions in forwarding strategies: \eg the time available to exchange data in each device encounter. 
\vspace{-0.1cm}
\subsubsection{Traffic analysis}
The participants were bound to only use the smarter-app for communications.
Thus only the services offered by the app generated traffic resulting in a total of 1,835 unique messages and 18,418 created bundles.
The mean of the messages was at 290.15 bytes with a standard deviation of 568.76 bytes.
Based on the interconnection times and the possible bandwidth of the WiFi channel, the generated traffic is well below the theoretical limit.
This is highly dependent on our design choice, to only offer text based services.
\vspace{-0.15cm}
\subsubsection{Messages}
Using the smarter-app each participant could sent and receive messages.
Those messages where then sent as a bundle via IBR-DTN.
Depending on the used service, the messages resulted in a unicast or multicast.
In total the participants generated 11,042 messages of which 1,348 where unicasts.
\begin{figure}[H]
	\centering
	\vspace{-0.25cm}
	\includegraphics[width=0.7\columnwidth]{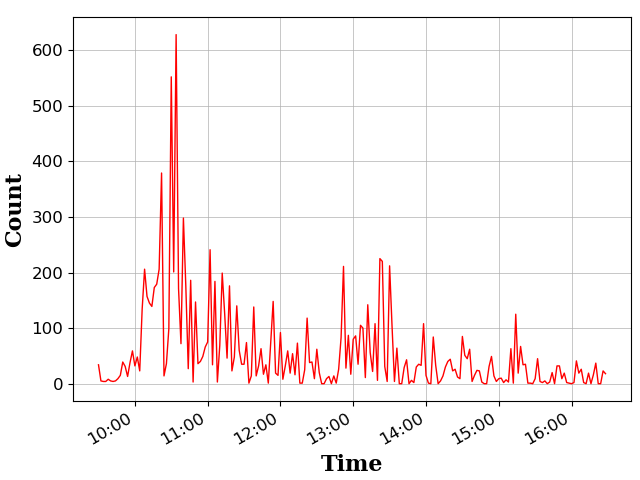}
	\vspace{-0.35cm}
	\caption{Received multicasts aggregated over 2 min.}
	\label{fig:broadcast_recv}
	\vspace{-0.25cm}
\end{figure}
As shown in Figure \ref{fig:broadcast_recv}, the participants started into the field test very motivated and created many messages during the first hour.
Resulting in a peak at around 10:30.
Afterwards the amount of messages slowly declined to almost none at around 12:30.
Upon the announcement of lunch and the start of the subscenarios the usage increased again.
The figure is extended one hour before and after the time frame we considered for our evaluation.
While we explicitly forbid to use the app before reaching the starting points most participants didn't comply with it.
For future field test we advise the enforce such rules directly in software.
	\vspace{-0.1cm}
\subsubsection{Cluster Coefficient}
A common metric to measure the interconnectivity of nodes over time is the cluster coefficient as described in \cite{watts1998collective}.
\begin{figure}[H]
	\centering
	\vspace{-0.25cm}
	\includegraphics[width=0.7\columnwidth]{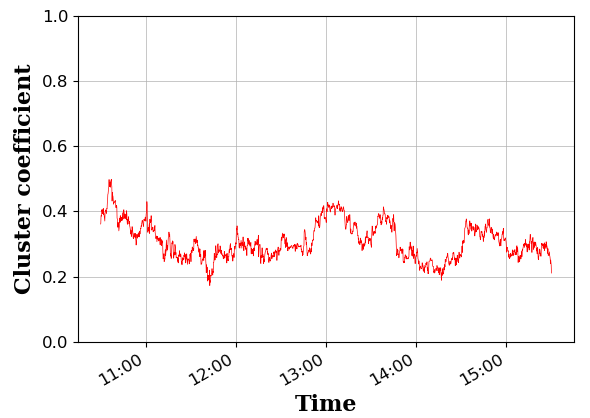}
	\vspace{-0.35cm}
	\caption{Cluster Coefficient}
	\label{fig:cluster_coef}
	\vspace{-0.35cm}
\end{figure}
The results in Figure \ref{fig:cluster_coef} show, that the highest connectivity was right at the beginning of the field test with around 0.41. 
This was expected, as the participants turned on their devices before the official start, while waiting to be brought to their starting point.
Two peaks at around 13:00 and 14:30 reflect the lunch break followed by our two subscenarios.
The low spot at 14:00 is not reflect in the GPS traces, meaning that the connectivity of the devices decreased while they should have been in close proximity.
This is most likely due to the then occurring rain and the reaction of the participants to seek shelter in buildings. 
The loss of \gls{los} and the walls of the buildings reduced the effective communication range.
	\vspace{-0.1cm}
\subsubsection{Propagation Delay}
An important metric in a \gls{dtn} is the propagation delay.
It describes the delay of a message from sender to destination.
\vspace{-0.20cm}
\begin{figure} [H]
	\centering
	\vspace{-0.25cm}
	\includegraphics[width=0.7\columnwidth]{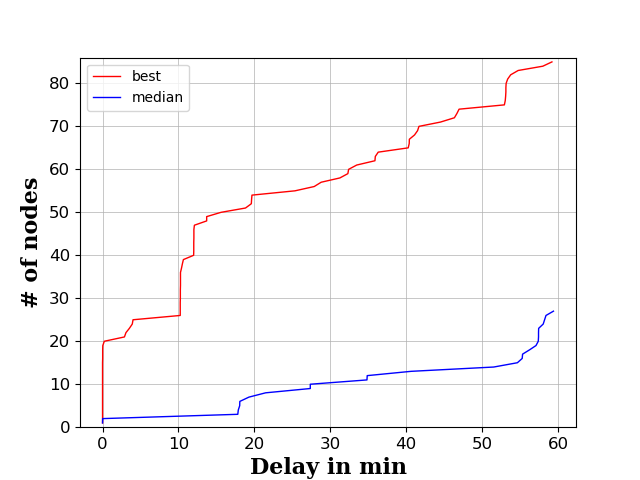}
	\vspace{-0.35cm}
	\caption{Propagation delay for multicasts}
	\vspace{-0.1cm}
	\label{fig:multicast_delay}
\end{figure}
\vspace{-0.2cm}
Figure \ref{fig:multicast_delay} shows the delay for the best performing multicast as well as for the median.
The figure is cut after 60 minutes, which was the defined lifetime of a bundle.
On average a bundle was successfully transmitted to 27 nodes or 21.77 percent of the network.
The best performing multicast reached a total of 86 nodes or 69.35 percent.
Overall the results show, that 20 percent of the messages got delivered to the destination directly.
This can be explained by looking at the mobility patterns of the participants.
Most of them formed groups, multicasts originating in one group reached each group member without delay.
Upon a meeting of groups, many messages are delivered in a short timeframe, which explains the steps visible in the figure.
The best performing multicast reached 20 nodes in under one second highlights the performance capabilities of the network.
Tests using WiFi Direct resulted in a maximum group size of 10, while decreasing stability.
If a message needs to be relayed the time it takes to reach its destination is uniform over its whole lifetime.
There is no clear evidence that the chances for a successfully delivery change over time. 
One reason is the storage capacity of the smartphones and our decision to not incorporate multimedia content.
As a result no message was dropped due to buffer size constraints, which would otherwise reduce the delivery chance over time.
%!TEX root = ../paper_traces.tex
	\vspace{-0.2cm}
\section{Conclusion}
\label{sec:conclusion}
In this paper, we presented a large-scale field test of a smart-\\phone-based ad hoc communication network in an emergency response scenario.
During a scripted emergency scenario, 125 participants used a mobile application to find family members, reach out for help, and share resources after a complete breakdown of the communication infrastructure.
We are the first to gather mobility traces, smartphone sensor data, application interaction patterns, and network logs of civilians in a large-scale field test specifically for emergency response.
We present a first analysis of the data gathered during the seven hour event, highlighting scenario-specific mobility and network characteristics.
Our results show, that a smartphone-based ad hoc network between more than one hundred smartphones provides sufficient connectivity for relevant emergency services.
Given the behavior of participants, connections lasted five minutes on average, exceeding the estimations stated in related work.  
Additionally, real-world impact of obstacles and crowd density lowered the achievable communication range.
Group-building contributed to these results, leading to devices having three neighbors on average.
Our results confirm the importance of real-world tests especially if systems are designed for scenarios that are heavily affected by human behavior.
We believe that our data contributes to the design and evaluation of works targeting disaster relief, especially when utilizing smartphone-based communication networks.
We are currently implementing simulation models based on our traces for The ONE~\cite{keranen2009one} as a starting point for further evaluation.
	\vspace{-0.15cm}
\begin{acks}
This work was supported in part by the BMBF within the SMARTER project, in part by the LOEWE initiative (Hessen, Germany) through the NICER project, and in part by the DFG as part of the CRC 1053 MAKI.
\end{acks}
	\vspace{-0.15cm}
\bibliographystyle{ACM-Reference-Format}
\balance
\bibliography{References}

\end{document}